\documentclass[12pt]{article}

\usepackage{amsmath}
\usepackage{amssymb} 
\usepackage{hhline}
\usepackage[scale={.75,.75}]{geometry}
\usepackage[T1]{fontenc}
\usepackage{bbm} 
\usepackage{mathrsfs} 
\usepackage{longtable}
\usepackage{caption}
\usepackage{cite}
\usepackage{epsfig}
\usepackage[multiple]{footmisc}

\newcommand{\GeV}{\ \text{GeV}}

\newcommand{\MeV}{\ \text{MeV}}
\newcommand{\s}{\ \text{s}}

\newcommand{\lstau}{ {\widetilde\tau_1}}
\newcommand{\mueff}{\tilde\mu}

\newcommand{\SO}[1]{\ensuremath{\text{SO}\!\left(#1\right)}}
\newcommand{\SU}[1]{\ensuremath{\text{SU}\!\left(#1\right)}}
\newcommand{\U}[1]{\ensuremath{\text{U}\!\left(#1\right)}}
\newcommand{\Z}[1]{\ensuremath{\mathbbm{Z}_{#1}}} 

\newcommand{\jvs}{\rule[-7pt]{0.00pt}{20pt}}

\numberwithin{equation}{section}
\numberwithin{table}{section}
\allowdisplaybreaks

\begin{document}
\date{\mbox{ }}

\title{ 
{\normalsize     
DESY 10-058\hfill\mbox{}\\
IPMU 10-0075\hfill\mbox{}\\}
\vspace{1cm}
\bf \boldmath Dynamical Matter-Parity Breaking\\
and Gravitino Dark Matter\\[8mm]}

\author{Jonas~Schmidt$^{1,2}$, Christoph~Weniger$^{1,2}$, Tsutomu~T.~Yanagida$^{2,3}$ \\[8mm]
{\normalsize \it $^1$ Deutsches Elektronen-Synchrotron (DESY),
 \it Notkestrasse 85, 22603 Hamburg, Germany}\\
{\normalsize \it$^2$ Institute for the Physics and Mathematics of the Universe (IPMU), 
 University of Tokyo,}\\
{\normalsize \it Chiba 277-8568, Japan}\\
{\normalsize \it$^3$ Department of Physics, University of Tokyo, Tokyo 113-0033, Japan}
}
\maketitle

\thispagestyle{empty}

\begin{abstract}
\noindent
Scenarios where gravitinos with GeV masses make up dark matter are known to be
in tension with high reheating temperatures, as required by $\textit{e.g.}$
thermal leptogenesis. This tension comes from the longevity of the NLSPs, which
can destroy the successful predictions of the standard primordial
nucleosynthesis. However, a small violation of matter parity can open new decay
channels for the NLSP, avoiding the BBN problems, while being compatible with
experimental cosmic-ray constraints. In this paper, we propose a model where
matter parity, which we assume to be embedded in the U$(1)_{B-L}$ gauge
symmetry, is broken dynamically in a hidden sector at low scales. This can
naturally explain the smallness of the matter parity breaking in the visible
sector. We discuss the dynamics of the corresponding pseudo Nambu--Goldstone
modes of $B-L$ breaking in the hidden sector, and we comment on typical
cosmic-ray and collider signatures in our model.
\end{abstract}

\newpage

\maketitle

\section{\label{sec:intro} Introduction}

Rapid proton decay in supersymmetric extensions of the standard model is
usually avoided by the introduction of $R$- (or matter-) parity. However, it
turns out that in scenarios with gravitino dark matter with masses in the
range $10$ to a few $100\GeV$, a small violation of this symmetry can be
desirable since this would reconcile thermal leptogenesis as a source for the
baryon asymmetry in the universe with standard primordial baryogenesis
\cite{bcx07} (for a recent exhaustive study of small bilinear $R$-parity
violation see \cite{bbx10}). The tight upper bounds on the corresponding
parity-violating couplings, stemming from laboratory experiments
\cite{add99,c05} and the non-observation of gamma-ray lines from gravitino
decays at Fermi LAT \cite{ax10}, then raise the question how the presence of
such small couplings can be understood theoretically. In this paper we
consider the possibility that the small violation of matter parity, which we
assume to be embedded into the $\U 1_{B-L}$ gauge symmetry, is linked to the
condensation of strongly interacting hidden sector fields at low energies.

Thermal leptogenesis \cite{fy86} via the out-of-equilibrium decay of heavy
right-handed neutrinos is one of the most promising models for the generation
of the baryon asymmetry in the universe. Furthermore, beside explaining the
observed excess of baryons over anti-baryons, the introduction of right-handed
neutrinos can also naturally explain the smallness of the neutrino masses;
this is known as the see-saw mechanism \cite{seesaw1,seesaw2}. Leptogenesis as
well as the see-saw mechanism favor very high masses for the right-handed
neutrinos, and it is well known that especially thermal leptogenesis requires
that the mass of the lightest right-handed neutrino should be larger than
$10^9\,$GeV~\cite{Davidson:2002qv}, which gives also a strong lower limit on
the reheating temperature.

In supersymmetric extensions of the standard model, gravitinos play a distinct
role due to their Planck-suppressed couplings to the observable world, and
they are an excellent candidate for being the dark matter observed in the
universe. It is well known that, given a high reheating temperature
$T_R\sim\mathcal{O}(10^9$--$10^{10}\GeV)$ and gravitino masses around
$m_{3/2}\sim\mathcal{O}(10$--$100\GeV)$, the thermal relic density of
gravitinos reproduces the correct dark matter abundance~\cite{Bolz:2000fu}, in
good agreement with thermal leptogenesis.

However, the above scenario is not free of problems: The next-to-lightest
supersymmetric particle (NLSP), which is often the stau, typically decays into
the gravitino with lifetimes that are of the order of minutes or days. This is
generically in conflict with the successful predictions of standard primordial
nucleosynthesis (BBN), since the NLSP decay releases large amounts of
electromagnetic or hadronic energy into the primordial plasma, destroying some
of the fragile elements and changing their abundance~\cite{Kawasaki:2004qu,
Jedamzik:2006xz}. Furthermore, if the NLSP is charged, it could form bound
states with $^4$He and $^8$Be, leading to a catalytic over-production of
$^6$Li and $^9$Be~\cite{Pospelov:2006sc, catalyzedBBN}.

Different solutions to the above problem were proposed: If in some special
scenarios the sneutrino or the stop is the NLSP, their late decay would not
affect the predictions of BBN strongly~\cite{Kanzaki:2006hm, DiazCruz:2007fc}.
In case of stau NLSPs, a large left-right mixing could lead to suppressed
relic abundances~\cite{left-right}. Entropy production between thermal
freeze-out of the NLSP and BBN could dilute the NLSP and gravitino abundance
enough to evade the BBN constraints~\cite{Pradler:2006hh, Buchmuller:2006tt}.
The NLSP mass could be nearly degenerate with the gravitino mass, which
reduces the released energy~\cite{Boubekeur:2010nt}. Or the NLSP could decay
into additional hidden sector states before conflicting with
BBN~\cite{DeSimone:2010tr, Cheung:2010qf}. Lastly, as mentioned above, in
cases where $R$-parity is weakly violated, the NLSP can decay into standard
model particles before the onset of BBN~\cite{bcx07}.\medskip

Theoretically, the existence of $R$-parity is not automatic. If it is a global
symmetry it is expected to suffer from quantum gravitational `wormhole'
effects \cite{h87,cl90}.  Therefore, it is necessary to assume an embedding
into a local symmetry, which is spontaneously broken at a high scale.
However, the requirement of anomaly cancellation strongly disfavors gauged $\U
1 _R$ proposals. Another possibility is to understand discrete
$R$-symmetry~\cite{kmy01} in the framework of higher-dimensional
constructions. The gauging then corresponds to local coordinate rotations in
the extra dimensions, and the $R$-parity of the effective theory is a relic of
the higher-dimensional Lorentz group. In string theory such discrete
$R$-symmetries may appear in specific compactifications~\cite{bs09}, however,
not generically.

Alternatively, proton stability in supersymmetric setups can be guaranteed by
matter parity.  This is natural in models which incorporate the see-saw
mechanism, since the required Majorana mass $M_S$ of the right-handed
neutrinos violates $B-L$ by two units, thereby breaking $\U 1_{B-L}$ to its
matter parity subgroup. Note that $\U 1_{B-L}$ is the only anomaly-free
Abelian extension of the standard model, and that it is canonically included
in $\SO{10}$ GUT proposals. Therefore, we consider the 'matter parity' to be
the $\Z2$ subgroup of the anomaly-free $\U1_{B-L}$ gauge symmetry throughout
this paper.

As argued above, the violation of $R$-parity or matter parity is bound to be
very small.  This can be achieved by \textit{e.g.}~connecting $R$-parity
breaking with $B-L$ breaking, via non-renormalizable operators~\cite{bcx07} or
at loop level in the context of string theory~\cite{knw09}.\medskip

In this paper, we discuss a scenario where the small matter-parity breaking is
related to the condensation scale $\Lambda$ of an asymptotically free hidden
sector gauge group factor. This scale can be naturally much lower than the $\U
1_{B-L}$ breaking scale $M_S$. If the hidden sector is charged under $B-L$, the
condensate breaks matter parity at low scale. It then dominantely couples to
the right-handed neutrinos, thereby inducing a sneutrino vacuum expectation
value and consequently small bilinear matter parity
violation~\cite{Barger:2008wn} of the order $\Lambda^2/ M_S$, which can lie in
the phenomenologically preferred regime.
\medskip

The paper is organized as follows: In the next section we will introduce the
hidden sector model and discuss its coupling to the visible sector; in section
3 we will discuss phenomenological aspects of the model, namely gravitino LSP
and stau NLSP decay, thermal production of hidden sector particles and their
decay; section 4 is devoted to further discussions and in section 5 we will
conclude.

\section{\label{sec:model} \boldmath A Model for Dynamical Matter-Parity
Breaking}

We consider an extension of the supersymmetric standard model with gauge group
\begin{align}
  \label{eq:G}
  G=\SU{3}_c\times \SU 2_L \times \U 1_Y \times \U 1_{B-L} \times \SU 2_{\rm
  hid} \ ,
\end{align}
where $\U 1_{B-L}$ will be broken spontaneously to matter parity at the high
scale $M_S$, and $\SU 2 _{\rm hid}$ is the gauge group in the hidden sector,
responsible for the low-energy matter-parity breaking.

We assume that light neutrino masses are generated in the standard way by the
see-saw mechanism after the breaking of $\U 1_{B-L}$ at a high scale $M_S$.
Large Majorana masses for the right-handed neutrinos $N^c$ are generated by a
singlet field $S$ in our model,
\begin{align}
  \label{eq:Wss}
  W_{\text{see-saw}} =&h^{(n)}_{ij} L_i N^c_j H_u+ \frac{1}{2}\lambda^S_{i} S
  N^c_i N^c_i \ ,
\end{align}
where we have taken the basis where the Majorana mass matrix for $N^c$ is
diagonal.  We assume $\langle S \rangle = M_S$. Then, the $N_i^c$ have
Majorana masses, $M_i=\lambda^S_i\times M_S$ for $i=1-3$, respectively, and we
define $M_1<M_2<M_3$ (we will assume that $M_3\simeq M_S$ below). For
successful thermal leptogenesis, we require $M_1> 10^9$ GeV. Since $\U
1_{B-L}$ is gauged initially, the corresponding Nambu--Goldstone (NG) mode
will be absorbed by the gauge field, which then decouples from the low-energy
effective theory.  The field $S$ necessarily has $B-L$ charge $-2$, hence a
global discrete $\Z 2$ subgroup remains unbroken during this
process\footnote{The generator of matter parity combines $B-L$ with
hypercharge, $P=(-)^{B-L+4 Y}$.}, the  `matter parity'.  We also assume the
presence of a field $\bar S$ with $B-L$ charge $+2$, required for anomaly
cancellation, without discussing the dynamics that give rise to the
condensation of $S$ and $\bar S$ in this paper. \medskip

The remaining matter-parity is then broken dynamically by the condensation of
quarks in the gauge group $\SU 2_{\rm hid}$ at the low scale $\Lambda \ll
M_S$.  The hidden sector  contains two doublet quarks $\mathcal Q^\alpha_1,
\mathcal Q^\alpha_2$ with $B-L$ charge $+1/2$, and two doublet quarks
$\mathcal Q^\alpha_3,\mathcal Q^\alpha_4$ with charge $-1/2$. Furthermore, we
require the existence of five neutral singlets $Z_{13}, Z_{14}, Z_{23},Z_{24}$
and $X$. Here and in the following $\alpha,\beta=1,2$ are $\SU 2_{\rm hid}$
indices. The low-energy degrees of freedom are the antisymmetric
combinations~\cite{s94}
\begin{align}
  V_{ij} &= - V_{ji} = \frac{1}{\Lambda} \mathcal Q_i^\alpha  \mathcal Q_{j
  \alpha} \ ,
\end{align} 
with convention $ \mathcal Q_{i \alpha} = \epsilon_{\alpha \beta} \mathcal
Q_i^{\beta}$, where $\epsilon_{\alpha \beta}$ is totally antisymmetric and
$\epsilon_{12}=1$.  

We consider the effective superpotential (cf.~\cite{s94,iy96})
\begin{align}
  \label{eq:Wdyn}
  W_{\rm dyn} = & \ X\left({\rm Pf} \left(V_{ij}\right)-\Lambda^2 \right)
  +\Lambda \left( Z_{13} V_{13} +Z_{14} V_{14}
  +Z_{23} V_{23}+Z_{24} V_{24} \right)\ ,
\end{align}
where ${\rm Pf} \left(V_{ij}\right)=V_{12}V_{34}+V_{14} V_{23} - V_{13}
V_{24}$ is the Pfaffian of the antisymmetric matrix $V_{ij}$ and the $X$ is
some composite state. The accidental symmetries of the model will be discussed
in Section~\ref{sec:disc}.
  
The superpotential~\eqref{eq:Wdyn} possesses a unique supersymmetric minimum
(which is in contrast to \cite{iy96}), with non-vanishing vacuum expectation
values for the two charged effective mesons,
\begin{align}
  \label{eq:Vvev}
  \langle V_{12} \rangle& = \langle V_{34} \rangle = \Lambda \ , &
  &\begin{array}{c}
    \langle V_{13} \rangle = \langle V_{14} \rangle = \langle V_{23} \rangle
    = \langle V_{24} \rangle = 0 \ ,\\ 
    \langle Z_{13} \rangle = \langle Z_{14} \rangle = \langle Z_{23} \rangle
    = \langle Z_{24} \rangle = \langle X \rangle = 0 \ .
  \end{array} 
\end{align}
Here, we assumed that $\langle V_{12} \rangle = \langle V_{34} \rangle$ is a
consequence of the soft mass terms for the $V_{ij}$, induced by SUSY breaking.
Since $V_{12}$ and $V_{34}$ have $B-L$ charge $+1$ and $-1$, respectively, we
conclude that matter parity is broken dynamically at the low scale $\Lambda$
in the hidden sector.\medskip

The only unsuppressed and renormalizable interaction in the superpotential,
connecting hidden and visible sector, is given by the term
\begin{equation}
  \label{eq:QQN}
  W\supset - f_i\mathcal Q^{\alpha}_3  \mathcal Q_{4 \alpha}N^c_i\;.
\end{equation}
Here, $f_i$ with $i=1,2,3$ are free parameters that we will fix below. In the
vacuum (\ref{eq:Vvev}) this becomes a linear term for the right-handed
neutrino multiplets, which together with the mass term in Eq.~(\ref{eq:Wss})
implies a non-vanishing vacuum expectation value for the corresponding
sneutrinos,
\begin{align}
  \label{eq:Ncvev}
  \langle N^c_i \rangle = \frac{f_i}{\lambda^S_i} \frac{\Lambda^2}{M_S} \ .
\end{align}
This vev generates matter parity violating couplings between the standard
model fields in the superpotential~\cite{Barger:2008wn}, which are of the
order
\begin{align}
  \nonumber
  W \supset&\ \mathcal{O}\left( \frac{\Lambda^2}{M_S} \right) L H_u
  +  \mathcal{O}\left( \frac{\Lambda^2}{M_S M_P} \right) L L E^c\\
  &+  \mathcal{O}\left( \frac{\Lambda^2}{M_S M_P} \right) L Q D^c
  +  \mathcal{O}\left( \frac{\Lambda^2}{M_S M_P} \right) D^c D^c U^c\ ,
  \label{eqn:afterCon}
\end{align}
coming from gauge invariant terms like $M_P^{-1} N^c LLE^c$ in the
superpotential.  Thus the matter-parity breaking is mostly bilinear in our
model, originating from the Yukawa couplings in \eqref{eq:Wss}. Its scale is
related to the condensation scale of the hidden sector gauge group.  The
effects on the matter-parity violation depend on many unknown Yukawa couplings
and we will assume a very simple situation where $f_3\leq1$ and $f_1=f_2=0$ in
this paper, to show the presence of a consistent parameter region in the
model. Note that we consider that the $N_1^c$ decay is relevant for
leptogenesis. Since $f_1=0$, the $N_1^c$ decays into the hidden sector are
suppressed and hence leptogenesis is not affected by the presence of the
hidden sector. Furthermore, the matter-parity violating couplings in
Eq.~\eqref{eqn:afterCon} are small enough to not wash-out the lepton asymmetry
again~\cite{Dreiner:1992vm} for the parameters we are interested
in~\cite{bcx07}. \bigskip

When the interactions (\ref{eq:QQN}) between hidden and visible sector are
neglected, we have a global $\U 1_{B-L}$ symmetry in the hidden sector, since
the corresponding gauge-symmetry is already broken by the condensation of $S$
and $\bar S$. This global $\U 1_{B-L}$ symmetry is then broken by $\mathcal
Q^{\alpha} \mathcal Q_{\alpha}$ condensations at the scale $\Lambda$,
producing a massless pseudo NG multiplet in the hidden sector. Through the
interactions~\eqref{eq:QQN}, this multiplet would receive a supersymmetric
mass $m_{\rm NG}=f_3^2\frac{\Lambda^2}{M_3} \sim
\mathcal{O}(\text{keV})$.\footnote{The true NG boson of the $B-L$ breaking is
a linear combination of this pseudo NG boson and the NG boson arises from the
high-energy $B-L$ breaking. This true NG boson is nothing but the one absorbed
in the $B-L$ gauge field.} In addition, there are hidden sector multiplets
with a mass $\sim\Lambda$.  However, after supersymmetry breaking, the soft
mass terms raise the masses of the modes in the pseudo NG multiplet. For the
pseudo NG boson this yields a mass (\textit{cf}.~app.~\ref{app:Xvev})
\begin{equation}
  \label{eqn:NGmass}
  m_{a}\sim f_3\sqrt{m_{3/2}\frac{\Lambda^2}{M_3}}
  \simeq 100 \MeV f_3 
  \left( \frac{m_{3/2}}{100\GeV} \right)^{\frac12}
  \left( \frac{\Lambda}{10^6 \GeV} \right)
  \left( \frac{M_3}{10^{16} \GeV} \right)^{-\frac12}
  \;,
\end{equation}
whereas the tree-level masses of the fermion partner $\psi$ and the radial
scalar pseudo NG component $\rho$ are given by
\begin{align}
  \label{eqn:NGmasses}
  m_\psi &\approx  m_{3/2}
  +\mathcal{O}\left(\frac{m_{3/2}^3}{\Lambda^2}\right)\ , &
  m_\rho &\approx 4 \ m_{3/2}\ ,
\end{align}
where we assumed that SUSY breaking is mediated to the hidden sector only via
gravity mediation. Below we will always assume that $m_\psi>m_{3/2}$, since we
are interested in gravitino dark matter.

\section{\label{sec:pheno} Phenomenology}
In this section we will discuss the decay of the gravitino and the NLSP, which
are general features induced by the matter-parity violation, as well as the
dynamics of the pseudo NG modes of $B-L$ breaking in the hidden sector, which
are special to the model we proposed. The resulting bounds on the parameter
space are summarized in Fig.~\ref{fig:Summary} for two reference scenarios.

\subsection{Gravitino and NLSP decay}
As stated above, the matter-parity violating effects on the MSSM sector are
mainly induced by the vev of $N_i^c$ and hence dominated by the bilinear term
\begin{equation}
  W\supset\mueff_i L_i H_u\;,
  \label{eqn:dominant}
\end{equation}
which is generated in the superpotential after condensation of the hidden
$\SU2_\text{hid}$ gauge group as described in the previous section. Here,
following Eqs.~\eqref{eq:Wss} and \eqref{eq:Ncvev}, the size of the
matter-parity violation is given by 
\begin{equation}
  \label{eqn:mutilde}
  \mueff_i\simeq f_3 h^{(n)}_{i3} \frac{\Lambda^2}{M_3}
  = 10^{-6}\GeV f_3\left( \frac{h^{(n)}_{i3}}{10^{-2}} \right)
  \left( \frac{\Lambda}{10^6\GeV} \right)^2
  \left( \frac{M_3}{10^{16}\GeV} \right)^{-1}\;,
\end{equation}
where we assumed that $f_3\leq1$ and $f_1=f_2=0$, as mentioned above.

For the phenomenological analysis, it is convenient to perform a rotation
between the higgs- and lepton-doublet superfields:
\begin{equation}
  H_d\to H_d - \varepsilon_i L_i\;, \hspace{2cm}
  L_i\to L_i + \varepsilon_i H_d\;,
  \label{eqn:rotation}
\end{equation}
where $\varepsilon_i\equiv \mueff_i/\mu$, and $\mu$ denotes the MSSM
$\mu$-term. The rotation eliminates the bilinear term in
Eq.~\eqref{eqn:dominant}, but it generates the matter-parity violating
trilinear terms $\frac12\lambda_{ijk}L_iL_jE_k^c$ and
$\lambda'_{ijk}L_iQ_jD_k^c$ with coefficients given by
$\lambda_{ijk}=-\varepsilon_{[i,}^{\phantom{(}} h^{(e)}_{j]k}$ and
$\lambda'_{ijk}=-\varepsilon_i^{\phantom{(}} h^{(d)}_{jk}$, respectively.  In
general, bilinear $R$-parity breaking terms such as \eqref{eqn:dominant} imply
a small but non-vanishing vev for the left-handed sneutrinos $\tilde\nu_i$. After the
rotation \eqref{eqn:rotation} it reads $\langle \tilde\nu_i\rangle\simeq
-\varepsilon_i v_d \left( 1+ \mu^2/ \widetilde m_{Li}^2 \right)$, where
$\widetilde m_{Li}^2$ denotes the soft slepton masses and soft $R$-parity
breaking terms were ignored\footnote{Vanishing soft bilinear $R$-parity
breaking terms can in fact only be demanded at one particular scale. In the
generic case without fine tuning bilinear $R$-parity breaking has 9
parameters, whose role for gravitino and NLSP decays was analyzed in detail in
\cite{bbx10}. Here we ignore phenomenological complications due to the
presence of soft bilinears, since in this paper we are mainly interested in
the theoretical origin of the supersymmetric term \eqref{eqn:dominant}.}.
Note that the operator $\frac12\lambda''_{ijk}U_i^cD_j^cD_k^c$, which finally
leads to proton decay, is not generated in this way.  Nevertheless, due to the
Planck-suppressed operator $M_P^{-1}N^cU^cD^cD^c$, it is generically present
in the superpotential with coefficients $\lambda'' \sim
\mathcal{O}(\Lambda^2/M_PM_3)$, \textit{cf.}~Eq.~\eqref{eqn:afterCon}.
Constraints from proton-stability require
$\lambda'_{11k}\lambda''_{11k}\lesssim 10^{-27}$~\cite{Dreiner:1997uz}, and a
sufficient condition to satisfy them is $\Lambda \lesssim 10^{6} \GeV$, where
we assumed that $M_3\sim10^{16}\GeV$ and $f_3=1$. This tight constrained will
be in general relaxed depending on the flavor structure of $\lambda_{ijk}''$
and $h_{ij}^{(d)}$ and on the value of $f_3$, and we will assume this to be
the case below.\footnote{Note that there is also the Planck-suppressed
operator $QQQL$ in the superpotential, which leads to proton decay and which
is not forbidden by the $B-L$ symmetry. We will assume that this operator is
suppressed by some unspecified mechanism, and note that it is not
significantly regenerated by breaking of matter-parity.}

Most importantly, the above violation of matter parity induces the decay of
the gravitino~\cite{bcx07}. The corresponding decay widths for the two-body
decay into $\gamma\nu$ is given by~\cite{Takayama:2000uz}
\begin{equation}
  \Gamma(\psi_{3/2}\to\gamma\nu)
  =\frac{1}{32\pi}|U_{\tilde\gamma\nu}|^2\frac{m_{3/2}^3}{M_P^2}\;,
  \label{eqn:GravitinoLifetime}
\end{equation}
where $U_{\tilde\gamma\nu}$ parameterizes the photino-neutrino mixing, and
$M_P=2.4\times10^{18}\GeV$ denotes the Planck-mass.  Assuming that SUSY
breaking only induces matter-parity conserving soft-masses at
tree-level\footnote{Note that the above rotation \eqref{eqn:rotation}
regenerates $B h_u h_d$, $\widetilde m_{Li}^2l^\ast_i l_i$ and $\widetilde
m_d^2 h_d^\ast h_d$, which then determine the photino-neutrino mixing angle
\cite{bbx10}.}, it can be approximated as \cite{bbx10}
\begin{equation}
  |U_{\tilde\gamma\nu}|\simeq \zeta v g\frac{s_W}{\sqrt{2}}
  \frac{M_{\tilde W}-M_{\tilde B}}{M_{\tilde W}M_{\tilde B}}
  \simeq 10^{-1} \zeta 
  \left( \frac{M_{\tilde B}}{150\GeV} \right)^{-1}
  \;,
  \label{eqn:MixingApprox}
\end{equation}
where $\zeta=\frac{v_d}{v}\sqrt{\sum \varepsilon_i^2}$ and $v_d= v \cos \beta
= \langle H_d^0\rangle$ denotes the higgs vev. We made use of the GUT relation
$M_{\tilde W}/M_{\tilde B}\simeq2$, where $M_{\tilde B}$ and $M_{\tilde W}$
are the bino and wino masses, respectively.  From this, we find for the
inverse decay width
\begin{equation}
  \Gamma(\psi_{3/2}\to\gamma\nu)^{-1}
  \simeq 4\times 10^{28}\s
  \left( \frac{\tilde\varepsilon}{10^{-8}} \right)^{-2}
  \left( \frac{M_{\tilde B}}{150\GeV} \right)^2
  \left( \frac{\tan\beta}{10} \right)^2
  \left( \frac{m_{3/2}}{100\GeV} \right)^{-3}\;,
  \label{eqn:Grav2body}
\end{equation}
with $\tilde\varepsilon\equiv \sqrt{\sum \varepsilon_i^2}\simeq
\varepsilon_3$, as long as $\varepsilon_1, \varepsilon_2\lesssim
\varepsilon_3$.  Other decay channels like $\psi_{3/2}\to W^\pm \ell^\mp$ and
$\psi_{3/2}\to Z^0\nu$, and decay channels with the hidden sector pseudo NG
boson in the final state, also contribute to the overall decay widths.
However, since the branching ratios are of comparable size~\cite{bcx07} for
the different two-body channels into SM particles, we will concentrate on the
decay into gamma-ray lines. This decay channel is constrained by gamma-ray
observations, which will give strong bounds on the matter-parity violating
parameter $\varepsilon_3$ in the region of interest.

The decay of the NLSP, which we take to be the lightest stau $\lstau$ for
definiteness in the rest of the paper, is mediated by the term $L_iL_jE_k^c$.
For simplicity we will assume that $\lstau$ is dominantly right-handed,
otherwise $L_iQ_jD_k^c$ would induce also decays into hadronic jets, see
Ref.~\cite{bcx07}. The corresponding decay width is dominated by decay into
$\nu_\mu\tau$ and $\nu_\tau\mu$ and given by
\begin{equation}
  \Gamma(\lstau\to\nu_\mu\tau, \nu_\tau\mu)
  \simeq2\times\frac{1}{16\pi}|\lambda_{233}|^2 m_{\lstau}\;,
  \label{eqn:StauDecay1}
\end{equation}
where we neglected fermion masses, and $\lambda_{233}=\varepsilon_3
h_{23}^{(e)} -\varepsilon_2 h_{33}^{(e)}\sim -0.1(\tan\beta/10)\
\varepsilon_2$. From this we obtain for the corresponding decay time
\begin{equation}
  \Gamma(\tilde\tau_1\to\nu_\mu\tau, \nu_\tau\mu)^{-1}\sim
  1\times10^{-7}\s\left(\frac{\varepsilon_2}{10^{-8}} \right)^{-2}
  \left( \frac{10}{\tan\beta} \right)^2
  \left( \frac{m_\lstau}{200\GeV} \right)^{-1}\;,
  \label{eqn:StauLifetime}
\end{equation}
from which it follows that the decay of the NLSP can easily happen before BBN.
Below we will always assume that $\varepsilon_2 \simeq \varepsilon_3 \simeq
\varepsilon$ for simplicity.\bigskip

\begin{figure}[t]
  \begin{center}
    \includegraphics[width=0.48\linewidth]{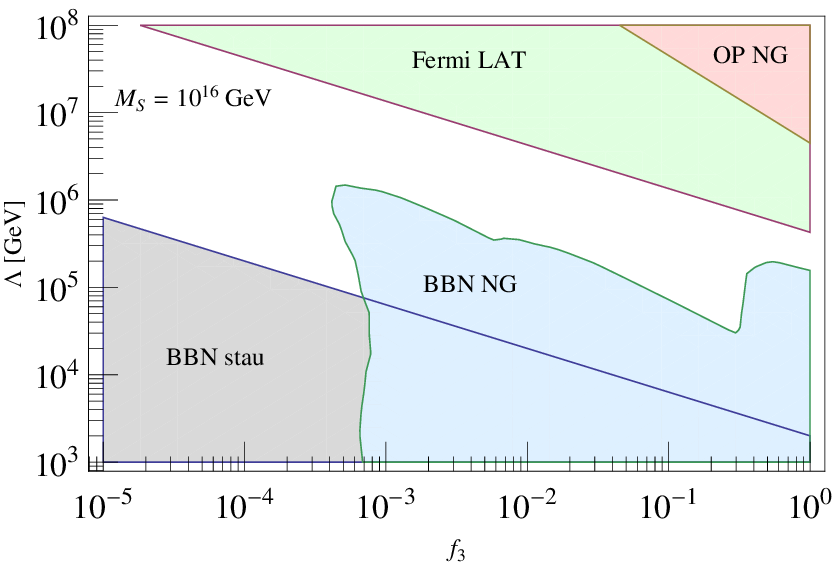}
    \includegraphics[width=0.48\linewidth]{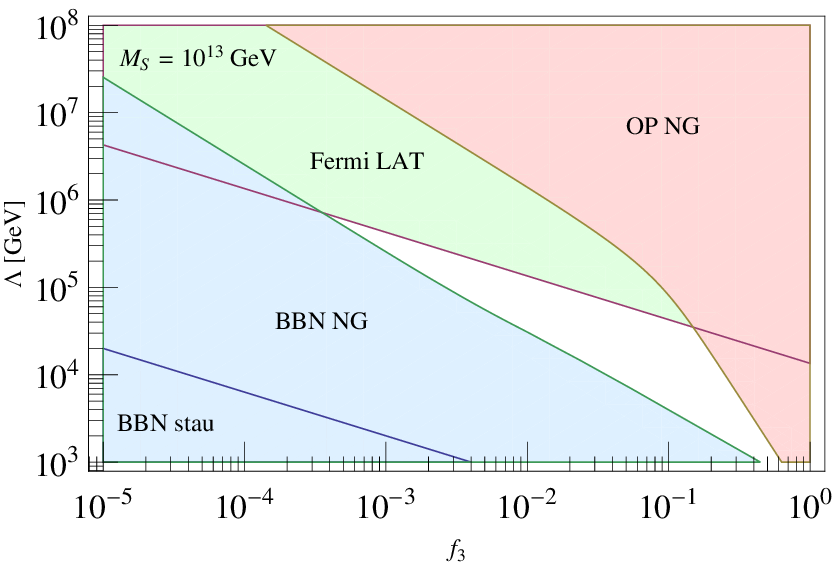}
  \end{center}
  \caption{Summary of constraints on condensation scale $\Lambda$, as function
  of the free parameter $f_3$, for two reference scenarios with
  $M_S=10^{16}\GeV$ (\textit{left panel}) and $M_S=10^{13}\GeV$ (\textit{right
  panel}). We assumed $m_{3/2}=120\GeV$, $y=10^{-1}$ and a reheating
  temperature of $T_R=10^{9.8}\GeV$, which yields the correct relic density
  for the thermally produced gravitino~\cite{Bolz:2000fu}. We show the bounds
  coming from catalytic production of $^6$Li during BBN by late decaying staus
  (BBN stau~\cite{catalyzedBBN}, for stau masses around 150--200 GeV), from
  the late decay of the fermionic partner of the pseudo NG boson of
  matter-parity breaking (BBN NG, from Ref.~\cite{Jedamzik:2006xz}, assuming
  branching ratio into hadrons $\simeq1$), and from the non-observation of
  gamma-ray lines from gravitino decay by Fermi LAT. For comparison, we show
  also the region where the pseudo NG boson relic density exceeds the
  observational limit $\Omega h^2 \simeq 0.11$ (OP NG). As apparent from the
  plots, bounds related to the pseudo NG modes become more sever for smaller
  values of $M_S$. We always assumed $f_1=f_2=0$.}
  \label{fig:Summary}
\end{figure}

As mentioned above, a number of astrophysical observations constrain the
lifetime of the stau and the gravitino, which translates into bounds on the
matter-parity breaking parameters. Firstly, the $\lstau$ has to decay before
BBN, with a lifetime shorter than $2\times 10^3\s$, to avoid catalytic
overproduction of $^6$Li~\cite{Pospelov:2006sc, catalyzedBBN}. This implies
the lower limit of roughly $\varepsilon \gtrsim 10^{-13}$. Furthermore, the
gravitino decay into gamma-ray lines is limited by observations of the Fermi
LAT satellite to lifetimes $\Gamma^{-1}_{\psi_{3/2}\to\gamma\nu}\gtrsim
10^{29}\s$, yielding an upper limit of $\tilde\varepsilon\lesssim
6\times10^{-9}(m_{3/2}/100\GeV)^{-3/2}$ in the mass range where the bounds
were evaluated, $60\GeV < m_{3/2} < 400\GeV$~\cite{Abdo:2010nc}. The
corresponding limits on the condensation scale $\Lambda$ are summarized in
Fig.~\ref{fig:Summary} for two reference scenario of parameters. In these
plots we also show the strong lower bounds on the scale $\Lambda$ and upper
bounds on $f_3$ that arise from the late decay of the pseudo NG modes of the
hidden sector, which we will discuss next.\bigskip

\subsection{Thermal production of hidden sector particles}

We will now discuss the cosmology of the hidden sector containing the
condensing $\text{SU}(2)_\text{hid}$ gauge group. Even when conservatively
assuming that reheating only affects the MSSM sector, hidden sector particles
are produced in the early universe by scattering of MSSM particles. This can
potentially lead to overclosure and can reintroduce problems with BBN.\medskip

From \eqref{eq:Wdyn} it follows that the hidden sector contains 10 chiral
multiplets with a mass at the condensation scale, $\sim\Lambda$, as well as
one chiral multiplet that remains light and is the pseudo NG mode of the
breaking of the accidental global $B-L$ symmetry in the hidden sector.  As
mentioned above, the pseudo NG boson $a$ acquires a small mass term,
suppressed by $M_S$, due to the coupling to the $B-L$ breaking visible sector,
see \eqref{eqn:NGmass}, whereas the fermionic partner $\psi$ and bosonic
partner $\rho$ of the pseudo NG boson get masses around the gravitino mass
$m_{3/2}$, see \eqref{eqn:NGmasses}. \bigskip

At temperatures above the condensation scale, $T\gg \Lambda$, the production
of hidden sector particles is dominated by the $M_S$-suppressed dim-5 operator
\begin{equation}
  W\supset\frac{y}{M_3}f_3\mathcal{Q}_3\mathcal{Q}_4 L H_u\;,
  \label{eqn:Dim5_1}
\end{equation}
which is generated after integrating out $N_3^c$, and by the $M_S^2$
suppressed dim-6 operators coming from integrating out the heavy $B-L$
vector-multiplet, which gives operators like
\begin{equation}
  K\supset \frac{(\mathcal{Q}_1^\dagger\mathcal{Q}_1 +
  \mathcal{Q}_2^\dagger\mathcal{Q}_2)
  L^\dagger L}{M_S^2}
  \label{eqn:dim6}
\end{equation}
in the Kahler potential $K$~\cite{Arvanitaki:2008hq}. Here we have defined
$y=\text{max}(h^{(n)}_{13},h^{(n)}_{23},h^{(n)}_{33})$, and we will ignore the
flavor structure for simplicity. Note that the term~\eqref{eqn:Dim5_1}
violates the $\U1 _\text{B-L}$ symmetry explicitly, but since this interaction
between hidden and visible sector is never in equilibrium, the lepton
asymmetry produced by leptogenesis is maintained.

In the first case, the matrix element for $2\to2$ scattering processes like
$lh_u\to \mathcal{Q}_3 \mathcal{Q}_4$ is of the order $|\mathcal{M}|^2\sim y^2
f_3^2 s/M_3^2$, where $s$ denotes the center-of-mass energy.  Following
Ref.~\cite{Hall:2009bx}, the yield $Y_{\mathcal{Q}_i}=n_{\mathcal{Q}_i}/S$ of
hidden sector quarks $\mathcal{Q}_i$, where $S$ is the entropy density, is
given by
\begin{equation}
  Y_{\mathcal{Q}_i}^{\text{UV},W}\sim y^2f_3^2\ \frac{T_R
  M_P}{\pi^7M_3^2 g_\ast^{3/2}} \simeq 2\times10^{-12}\ y^2 f_3^2
  \left( \frac{T_R}{10^9\GeV} \right)
  \left( \frac{M_3}{10^{16}\GeV} \right)^{-2}\;,
  \label{eqn:Yield1}
\end{equation}
where we have used $g_\ast\simeq 915/4$ as the number of effective
relativistic degrees of freedom of the MSSM.  Similarly, in the second case,
the operator in \eqref{eqn:dim6} induces $2\to2$ scattering processes like
$l^\ast l\to \mathcal{Q}_1^\ast \mathcal{Q}_1$, and the corresponding matrix
element is of the order $|\mathcal{M}|^2 \sim s^2/M_S^4$.  The corresponding
yield is given by
\begin{equation}
  Y_{\mathcal{Q}_i}^{\text{UV},K}\sim \frac{T_R^3 M_P}{\pi^7M_S^4
  g_\ast^{3/2}}
  \sim 10^{-26}\left( \frac{T_R}{10^9\GeV} \right)^3
  \left( \frac{M_S}{10^{16}\GeV} \right)^{-4}\;.
  \label{eqn:Yield1}
\end{equation}

The above processes are most efficient at the highest temperature, $T_R$. The
energy density produced in the hidden sector can be hence estimated to be
$\rho_\text{HS}\sim T_R n_\text{HS}$, where $n_\text{HS}=
Y^\text{UV}_{\mathcal{Q}_i} S$ denotes the particle density in the hidden
sector shortly after reheating. Assuming that particles in the hidden sector
quickly thermalize, this corresponds to a temperature $T_\text{HS}$ in the
hidden sector that is given by $(\pi^2/30) g_\text{HS}^{(R)} T_\text{HS}^4 =
\rho_\text{HS}$~\cite{Kolb:1981hk}. From this it follows that the temperature
in the hidden sector shortly after reheating is $T_\text{HS} \sim
(g_\text{VS}^{(R)}/g_\text{HS}^{(R)})^{1/4} (Y_{\mathcal{Q}_i}^\text{UV}
)^{1/4} T_R$, where $g_\text{HS}^{(R)}$ and $g_\text{VS}^{(R)}$ denote the
corresponding effective degrees of freedom in the hidden and visible sector at
high energies.

Depending on the details of the symmetries of the hidden sector, some of the
heavy hidden sector states with masses around the condensation scale
$\sim\Lambda$ can be stable, which could lead to overclosure of the universe.
However, despite their large masses, which can exceed the unitarity bound
$\Lambda \lesssim 10^5\GeV$~\cite{Griest:1989wd}, sufficiently strong
annihilation of the hidden sector states is in principle possible due to the
large temperature difference between hidden and visible sector, which reduces
the relic density by a factor of order $T_\text{HS}/T_\text{VS}\sim
(Y_{\mathcal{Q}_i}^\text{UV})^\frac14$, see Ref.~\cite{Feng:2008mu}.  However,
a detailed study of this annihilation process is beyond the scope of this
paper. Subsequently we will assume for simplicity that all these potentially
stable heavy hidden sector states efficiently annihilate into the pseudo NG
modes, and we will consider the subsequent effects of these NG modes only.
Alternatively, depending on the underlying symmetries, the heavy hidden sector
particles could directly decay into Standard Model particles before BBN, thus
avoiding the overclosure problems from the beginning. We will discuss this
situation shortly in section~\ref{sec:disc} below.

To estimate the hidden sector temperature and density of pseudo NG modes at
later times, one can make use of the fact that the entropy densities, which is
given by $S_\text{HS/VS}\simeq (2\pi^2/45) g_\text{HS/VS}T_\text{HS/VS}^3$, in
the hidden and visible sector are conserved separately~\cite{Kolb:1981hk}.
>From this it follows that today $T_\text{HS}\sim
(g_\text{VS}/g_\text{HS})^{1/3} (Y^\text{UV}_{\mathcal{Q}_i})^{1/4}
T_\text{VS}$, where $g_\text{VS}$ and $g_\text{HS}$ denote the corresponding
effective degrees of freedom of the hidden and visible sector, now at low
energies.  Noting that the particle density today is given by
$n_\text{HS}\approx g_\text{HS}/\pi^2 T_\text{HS}^3$, this gives an effective
particle yield of $Y_\text{NG}^\text{UV} = n_\text{HS}/S_\text{VS}\sim 0.2
(Y^\text{UV}_{\mathcal{Q}_i})^{3/4}$.  The corresponding relic density of the
different pseudo NG modes is given by
\begin{equation}
  \Omega_{a,\rho,\psi}^{\text{UV}}h^2=
  2.74\times10^{10}
  \left( \frac{m_{a,\rho,\psi}}{100\GeV} \right)Y_{a,\rho,\psi}^\text{UV}\;,
  \label{eqn:Relic1}
\end{equation}
and we take for definiteness
$4Y_{a}^\text{UV}=4Y_\rho^\text{UV}=2Y_\psi^\text{UV} =
Y_\text{NG}^\text{UV}$.\medskip

At temperatures below the condensation scale, $100\GeV\lesssim T\ll\Lambda$,
hidden sector particles are predominantly produced via the renormalizable
operator
\begin{equation}
  W\supset yf_3\frac{\Lambda}{M_3}V_{34}L H_u\;.
  \label{eqn:W2}
\end{equation}
Here, the decay of particles in the thermal bath into the hidden sector
particles is relevant, like \textit{e.g.}~decays of the higgs~$h^0\to\psi\nu$.
The corresponding relic density is given by~\cite{Hall:2009bx}
\begin{equation}
  \Omega_{a,\rho,\psi}^{IR} h^2 \approx \frac{10^{27}}{g_\ast^{3/2}}
  m_{a,\rho,\psi}\sum_j\frac{\Gamma_j}{m_j^2}\sim 10^2\ 
  f^2_3y^2
  \left( \frac{m_{a,\rho,\psi}}{100\GeV} \right)
  \left( \frac{\Lambda}{10^6\GeV} \right)^2
  \left( \frac{M_3}{10^{16}\GeV} \right)^{-2}
  \;,
  \label{eqn:Relic3}
\end{equation} 
where $\Gamma_j\sim 1/8\pi (yf_3\Lambda/M_3)^2 m_j$ denotes the approximate
decay widths of particles $j$ in the bath into the hidden sector modes, and we
used $g_\ast\approx100$.  This can dominate over \eqref{eqn:Relic1} when
$\Lambda$ is large.\medskip

\subsection{Decay of the hidden sector pseudo NG modes}
Finally, we discuss the decay of the pseudo NG modes. The fermionic pseudo NG
boson partner $\psi$ mixes with the neutrinos $\nu_i$ and inherits their SM
interactions.  This mixing is induced by the term~\eqref{eq:QQN}, and given by
(for details see app.~\ref{app:massmat})
\begin{equation}
  \theta_i\simeq \frac{h^{(n)}_{i3}f_3}{\sqrt2} 
  \frac{v_u \Lambda}{M_3 m_\psi}\;,
  \label{eqn:Mixing}
\end{equation}
which is suppressed by the $B-L$ breaking scale $M_S$. Here, the index $i$
refers to neutrino flavor eigenstates. Using \eqref{eqn:mutilde}, the size of
the mixing angle can be expressed in term of the matter-parity breaking
parameter like
\begin{equation}
  \theta_i\simeq 1.2\times10^{-13}
  \left( \frac{\mueff_i}{10^{-7}\GeV} \right) \left(
  \frac{m_\psi}{100\GeV} \right)^{-1}
  \left(\frac{\Lambda}{10^6\GeV}  \right)^{-1}
  \;.
  \label{eqn:eq2}
\end{equation}
If $m_\psi>M_Z$, the two-body decay $\psi\to Z^0\nu_i$ is kinematically
allowed, and the corresponding decay width is given by
\begin{align}
  \Gamma(\psi\to Z^0\nu_i)&\simeq\nonumber
  \frac{e^2}{16\pi}\frac{\theta^2_i}{8s_W^2c_W^2}
  \frac{m_\psi^3}{M_Z^2}
  \left( 1+\frac{M_Z^2}{m_\psi^2}-2 \frac{M_Z^4}{m_\psi^4} \right)
  \left( 1-\frac{M_Z^2}{m_\psi^2} \right)\\
  &\simeq
  (0.03\,\text{s})^{-1}\left( \frac{\theta_i}{10^{-11}} \right)^2
  \left( \frac{m_\psi}{150\GeV} \right)
  \frac{m_\psi^2}{M_Z^2}
  \left( 1+\frac{M_Z^2}{m_\psi^2}-2 \frac{M_Z^4}{m_\psi^4} \right)
  \left( 1-\frac{M_Z^2}{m_\psi^2} \right)\;.
  \label{eqn:Lifetime}
\end{align}
For masses around $m_\psi\sim150\GeV$ and mixing angles $\theta_i\sim10^{-11}$
this gives inverse decay widths of the order of $\tau\sim 0.03\s$, which is
well before and save for BBN~\cite{Jedamzik:2006xz}. However, when $\theta_i$
is smaller, BBN bounds become very relevant.\footnote{Note that the decay into
gravitinos, $\psi\to a \psi_{3/2}$, is typically very slow with lifetimes of
the order of $\tau\gtrsim 10^8\s$ or larger, and can be safely neglected.} The
lifetime of $\psi$ is further reduced by the decay channels $\psi\to
W^\pm\ell^\mp$, whose decay width is obtained from \eqref{eqn:Lifetime} by
$\Gamma_{\psi\to W^\pm \ell_i^\mp}=4 c_W^2 \Gamma_{\psi\to Z^0\nu_i}$ and
substituting $M_{Z}\to M_{W^\pm}$. Note, that if $m_\psi<M_{W^\pm}$, the much
slower three-body decays like $\psi\to e^+ e^- \nu$ would become important.
They would give rise to much stronger lower bounds on the mixing angle
$\theta$ than in the regime where two-body decay is allowed.\medskip

The bosonic pseudo NG boson partner $\rho$ can perform the two-body decays
like $\rho\to\tilde h_u \nu$, if kinematically allowed, as follows from the
effective operator \eqref{eqn:Dim5_1}. This requires that $M_{\tilde h_u}<
m_\rho$.  The lifetime of $\rho$ is typically much smaller than the lifetime
of $\psi$, since its decay is not further suppressed by gauge couplings. When
the above decay is kinematically forbidden, three-body decays like $\rho\to
b\bar b\nu$ become relevant. However, this depends on the details of the MSSM
spectrum, and we will conservatively assume that the decay of $\rho$ is always
faster than the decay of $\psi$.

The pseudo NG boson $a$ in principle also decays into SM particles. The
interactions with the visible sector are at low energies mainly mediated by
the operators $V_{34}L_i H_u$, and therefore the interaction is suppressed
only by a factor $h^{(n)}_{i3}\Lambda/M_3$. However, since the particle
interacts with the visible sector only via a matter parity violating operator,
final states for its decay can only be odd under matter parity. But no matter
parity odd final states with integer spin number exists below the mass of the
lightest gaugino, and hence the particle is essentially stable and one has to
require that $\Omega_a^\text{IR,UV} h^2\lesssim0.1$ to avoid
overclosure.\footnote{In fact, the decay rate is doubly suppressed by the
small matter-parity violation, and loop suppressed since the decay involves
the higgs-multiplets, \textit{e.g.}~$\Gamma(a\to\gamma\gamma)\sim \alpha^2(y
f_3 \Lambda/M_3\ \mueff /\mu)^2\ m_a^5/\mu^4\sim
\mathcal{O}(10^{42}\s)^{-1}$.}\medskip

Lastly, we note that the invisible decay of gravitinos into the pseudo NG
boson and neutrinos, which is kinematically allowed, happens with lifetimes
around $\tau\sim10^{32}\s\ (10^{-12}/\theta)^2$ for $\sim100\GeV$ masses, and
can hence be neglected. Furthermore, the decay $\tilde\tau\to\tau \psi$, which
is also in principle possible, is double suppressed by the small matter-parity
breaking $\tilde\mu/\mu$, and hence negligible with respect to the decay mode
$\tilde\tau\to\tau\nu$.\bigskip

The bounds on the parameter space of our model are summarized in
Fig.~\ref{fig:Summary} for two reference scenarios. We show bounds from the
non-observation of gamma-ray lines with Fermi LAT, where we have assumed that
gravitinos make essentially all of dark matter today.  Furthermore, we show
bounds from BBN that stem from the late decay of the stau via the induced
R-parity violation, and bounds from BBN that come from the late decay of the
fermionic and bosonic pseudo NG boson partner. Lastly, we also show the region
where the relic density of the pseudo NG boson would exceed the observational
limit.

\section{Discussion}
\label{sec:disc}

As shown in Fig.~\ref{fig:Summary}, cosmologically viable models with high
reheating temperatures and gravitino dark matter can be found for large enough
$\U1_{B-L}$ breaking scales $M_S\gtrsim\mathcal{O}(10^{13}\GeV)$ and
$\SU2_\text{hid}$ condensation scales $\Lambda \sim \mathcal{O}
(10^3$--$10^8\GeV)$.  We also assumed suppressed couplings between hidden
sector and the lighter two right-handed neutrinos, adopting $f_1=f_2=0$ for
simplicity. In these models, the communication between the matter-parity
breaking $\SU2_{hid}$ sector and the visible sector is suppressed by the mass
of the heaviest right-handed neutrino, $M_3\simeq M_S$. As an important
consequence, the mass of the pseudo NG boson of the breaking of the accidental
global $B-L$ symmetry in the hidden sector, as well as its relic number
density, are small enough to yield a relic density in agreement with the
observations, so that the pseudo NG boson constitutes only a subdominant
component of dark matter. Furthermore, the relic density and the lifetime of
the pseudo NG boson partners can be small enough to be in agreement with the
tight BBN constraints. Note that requiring that the lighter two right-handed
neutrinos do not change the above phenomenology implies upper limits on $f_1$
and $f_2$ that are at least of the order $f_i\lesssim f_3 M_i / M_3$, since
the interactions between hidden and visible sector depends on the ratio $f_i /
M_i$.

Due to the strong BBN bounds on the relic abundance and the lifetime of the
pseudo NG modes of $B-L$ breaking, large values of matter-parity breaking are
favored in our model.  This implies both a lifetime of the gravitino closer to
the experimental limit, as well as shorter decay lengths of the NLSPs. For
values of matter-parity breaking close to the Fermi LAT limit, we estimate the
decay length of the stau to be of the order of $\mathcal{O}(100\ \text{m})$.
It is very interesting, that this is in principle accessible by the
LHC~\cite{Asai:2009ka}. However, the quantitative details depend
\textit{e.g.}~on the flavor structure of the stau couplings, and we refer to
Ref.~\cite{bbx10} for a detailed discussion.\bigskip

In general, the low-scale breaking of discrete symmetries can induce serious
cosmological domain wall problems, which arise after the completion of
inflation and thus are never diluted~\cite{Zeldovich:1974uw}. In the presented
scenario these problems can be avoided in two different ways.

Firstly, estimating the maximum temperature $T_\text{HS}$ of the hidden sector
immediately after reheating, by equating $\rho_\text{HS}\sim
T_\text{VS}n_\text{VS} Y \sim T_\text{HS}^4$, shows that the hidden sector
temperature after inflation does never exceed the breaking scale $\Lambda$ in
a large part of the allowed parameter space in Fig.~\ref{fig:Summary} (namely
for $f_3\lesssim0.1$ in case of $M_S=10^{16}\GeV$). In these cases, the
matter-parity symmetry is never restored after inflation, despite its low
breaking scale, thus avoiding all domain wall problems.  

Secondly, even if the symmetry is restored, domain wall problems can be in
principle avoided if the discrete symmetry is embedded into an $\U 1$ gauge
symmetry, since in that case the minima of the potential are continuously
connected and the domain walls are no longer stable. This mechanism requires
however that the gauge symmetry is broken after inflation, since only then the
domain walls are bounded by cosmic strings and they shrink and decay rapidly
when the $\U 1$ is spontaneously broken after inflation
ends~\cite{Preskill:1991kd}.  Therefore, in cases where the $\U1_{B-L}$
breaking occurs after inflation, our model is free of cosmological domain wall
issues, despite the low scale breaking. \bigskip

The superpotential (\ref{eq:Wdyn}) has four global $\U 1$ symmetries, with
charge assignments as listed in Table~\ref{tab:dynU}.
\begin{table}
  \centering
  \begin{tabular}{c|cccc|cccc|cc}
    \jvs  & $\mathcal Q^\alpha_1$&$\mathcal Q^\alpha_2$& 
    $\mathcal Q^\alpha_3$&$ \mathcal Q^\alpha_4$& 
    $Z_{13}$&$Z_{14}$&$Z_{23}$&$Z_{24}$&$X$ \\
    \hline 
    \hline
    \jvs $\U 1_{B-L}$ & 
    $\frac 12$& $\frac 12$&$ -\frac 12$&$ -\frac 12$
    & 0&0&0&0&0\\
    \hline
    \jvs $\U 1_a^{\rm(acc.)}$  & $\frac 12$ & $-\frac 12$ & $\frac 12$ & $-\frac 12$
    & $-1$ & 0 & 0 & 1 & 0  \\
    \jvs $\U 1_b^{\rm(acc.)}$  & $\frac 12$ & $-\frac 12$ & $-\frac 12$ & $\frac 12$
    & 0 & $-1$ & 1 & 0 & 0  \\
    \jvs $\U 1_R^{\rm(acc.)}$  & $-\frac 12$& $-\frac 12$&$ \frac 12$&$ \frac 12$
    & 2 & 2 & 2 & 2 & 2  \\
  \end{tabular}
  \caption{\label{tab:dynU}The global symmetries of the superpotential
  (\ref{eq:Wdyn}).  The standard model fields  are not charged under $\U
  1_{a,b}^{\rm(acc.)}$, whereas their charges with respect to $\U 1_{B-L}$ and
  $\U 1_{R}^{\rm(acc.)}$ are the canonical ones. The fields $S, \bar S$ are
  neutral with respect to the accidental symmetries.}
\end{table}
Besides $\U 1_{B-L}$ it exhibits three additional accidental symmetries, $\U
1_{a}^{\rm (acc.)}, \U 1_{b}^{\rm (acc.)}$ and the $R$-symmetry $\U 1_R^{\rm
(acc.)}$.  They reflect the absence of linear terms and quadratic terms for
the fields $Z_{ij}$ in the superpotential. These symmetries are either
explicitly broken by higher order operators, for example trilinear terms which
do not influence the vacuum (\ref{eq:Vvev}), or they are symmetries of the
full Lagrangian. In the latter case one may gauge the symmetries, since they
are anomaly free.  However, we do not want to have additional massless gauge
fields in the model, and hence we rather consider unbroken discrete subgroups
of these symmetries in that case.   \medskip

The accidental $R$-symmetry disappears if one introduces mass terms like $W=M
Z_{13}Z_{24}$ in the superpotential. Such terms are compatible\footnote{In
addition new supersymmetric minima with $\langle X\rangle =\pm \Lambda^2/M$
occur, where $B-L$ remains unbroken.} with the vacuum under consideration.
However, if one wants to avoid the prediction of light fermions with masses
$\sim \Lambda^2/M$, generated by a see-saw like mechanism in the hidden
sector, we need that $M \ll\Lambda$. This may follow if the explicit violation
of $\U 1_{R}^{\rm(acc.)}$ occurs only via higher-dimensional operators in the
hidden sector.  Alternatively, one may consider that the fields $Z_{ij}$ are
moduli of an underlying theory, and therefore have no mass terms. Finally,
note that the combination $R^{\rm acc.}+B-L$ is orthogonal to the vevs
$\langle V_{12} \rangle, \langle V_{34} \rangle$. After supersymmetry breaking
this symmetry is broken to its $\Z 2$ subgroup, which is trivial up to
hypercharge.\medskip

The absence of linear terms $W=C_{ij}Z_{ij}$ is crucial for our model, since
such terms completely remove the vacuum (\ref{eq:Vvev}). It can be explained
by promoting a discrete subgroup of $\U 1^{\rm acc.}_{a,b}$ to a rigorous
symmetry of the hidden sector. An important consequence is then that the
lowest mass states of $V_{13}, V_{14}, V_{23}$ and $V_{24}$ become stable
together with $Z_{ij}$ due to the $\U1_{a,b}^{(\rm acc.)}$ symmetry. As
discussed above, this can be problematic since their relic mass density may
overclose the universe, if annihilation into the pseudo NG multiplet is not
efficient enough.  In this cases it would be better to break the $\U1_{a,b}$
symmetries explicitly, to allow the decay of the hidden sector particles. As
an example, we note that the linear combination $\U 1_{a+b}$ alone is needed
in order to forbid the linear terms in $Z_{ij}$. This combination is preserved
if we enlarge (\ref{eq:Wdyn}) to
\begin{align}
  \label{eq:Wdyn2}
  W_{\rm dyn} = & \ X\left({\rm Pf} \left(V_{ij}\right)-\Lambda^2 \right)
  +\Lambda V_{13} \left(Z_{13}+Z_{14} \right)
  + \ldots+\Lambda V_{24} \left(Z_{23}+Z_{24} \right) \ .
\end{align}
We can then gauge $\U 1_{a+b}$ and assume it is spontaneously broken to a $\Z
2$ subgroup at a high scale $\Lambda'$, such that the heavy hidden sector
states have odd parity. If we now identify this $\Z 2$ with the Peccei--Quinn
like $\Z 2$ of the SSM (cf.~\cite{bs09}, in $\SU 5$ notation)\footnote{This $\Z 2$ forbids the
dangerous superpotential operators ${\bf 10} \ {\bf 10} \  {\bf 10} \ {\bf
5^*}, {\bf 10} \ {\bf 5^*} \ {\bf 5^*}$, but not ${\bf 5^*} H_u$.}
\begin{align}
  P\left( {\bf 10} \right) &=P\left( H_d\right) =- \ , &
  P\left( {\bf 5^*}\right) &=P\left( H_u \right)=P\left(N^c \right)=+ \ ,
\end{align}
decay channels of the hidden sector states to $H_u H_d$ open up. This can
resolve the overclosing issue of the hidden sector from the start. Since the
breaking of this $\Z 2$ is then related to the generation of the $\mu$ term,
we expect that also the linear terms in $Z_{ij}$ remain small, since they are
additionally suppressed by the scale $\Lambda'$.

\section{Conclusions} 

In this paper we have discussed a possible mechanism that generates a small
matter-parity violation in the visible sector. Such a violation of
matter-parity can resolve the tension between gravitino LSPs and high
reheating temperatures as required by thermal leptogenesis. In our scenario,
visible sector and part of the hidden sector are simultaneously charged under
a gauged $\U1_{B-L}$. This $\U1_{B-L}$ is broken at a high scale $M_S$ by
right-handed neutrino masses to its matter-parity $\Z2$ subgroup.
Matter-parity is then subsequently broken completely in the hidden sector by a
$\SU2_\text{hid}$ quark condensate at a scale $\Lambda$. This mechanism
induces a small vev $\sim \Lambda^2/M_S$ for the right-handed neutrinos,
giving rise to bilinear matter-parity breaking in the visible sector. Due to
matter-parity breaking, the NLSP, typically a stau or neutralino, can then
decay into standard model particles before conflicting with BBN. 

Since the hidden sector is almost separated from the high-energy $B-L$
breaking sector, the hidden sector possesses an accidental global $B-L$
symmetry which is explicitly broken only through the interactions with the
right-handed neutrinos. The effects of the explicit breaking are suppressed by
$1/M_S$ at low energies, and hence the pseudo NG modes arising from the $B-L$
breaking in the hidden sector acquire only very small masses. The tightest
constraints on our proposed model come from the dynamics of these pseudo NG
modes. The pseudo NG boson itself acquires $\mathcal{O}($MeV--GeV$)$ masses,
remains practically stable, and can be easily overproduced if the coupling
between hidden and visible sector is too large. The fermionic and bosonic
partners of the pseudo NG boson, however, acquire $\mathcal{O}(100\GeV)$
masses, decay early in the universe and can reintroduce problems with BBN. We
showed that the model can be phenomenologically viable for large enough $B-L$
breaking scale $M_S\gtrsim\mathcal{O}(10^{13}\GeV)$, and for condensation
scales $\Lambda\sim\mathcal{O}(10^3$--$10^8\GeV)$, when the coupling to the
lighter right-handed neutrinos is suppressed.

An important consequence of gravitino dark matter scenarios with small
breaking of matter-parity is that the gravitino decays, producing gamma-ray
lines which are potentially observable with instruments like Fermi LAT. We
found that within our model, large values of the matter-parity violation are
favored, since this reduces the lifetime of the problematic pseudo NG modes.
This suggests lifetimes of the gravitino close to the experimental limit, as
well as short decay lengths of the NLSP. For a stau NLSP, we found a lower
limit for typical stau decay lengths that is given by $\gtrsim\mathcal{O}(100\
\text{m})$. Such events are in principle detectable at the LHC.

\section*{Acknowledgements}
We would like to thank Wilfried Buchmuller and Jan Hajer for useful
discussions. JS and CW would like to thank the IPMU for kind hospitality during
the main stages of this work. This work was supported by the World Premier
International Research Center Initiative (WPI Initiative), MEXT, Japan.

\appendix

\section{\label{app:massmat} Fermion Mixing Matrix}

The superpotential contributions \eqref{eq:Wss}, \eqref{eq:Wdyn}
and~\eqref{eq:QQN} imply mixings between the 13 neutral fermions 
\begin{align} 
  \tilde B, \tilde W, \tilde h_u^0, \tilde h_d^0, \nu_i,
  N^c_i, \tilde V_{12}, \tilde V_{34}, \tilde X \ , 
\end{align} 
where the first four fields are the MSSM gauginos and neutral higgsinos,
$\nu_i$ and $N^c_i$ are the left- and right-handed neutrinos, respectively,
and the last three fields are fermions of hidden sector chiral multiplets. In
this basis the mass matrix  is to leading order (and up to order one
coefficients)
\begin{align}
  \label{eq:M1}
  \hat M &=
  \left( \begin{array}{cccc|cc|ccc}
    \jvs M_B & 0 & M_Z   & -   M_Z & \alpha_i \Lambda^2/M_S & 0 & 0 & 0 & 0 \\
    \jvs 0 & M_W & -M_Z    &   M_Z  & \alpha_i  \Lambda^2/M_S & 0 & 0 & 0 & 0 \\
    \jvs    M_Z  & -  M_Z  & 0 & -\mu &   \beta_i \Lambda^2/M_S& \alpha_i \Lambda^2/M_S & 0 & 0 & 0  \\
    \jvs  -   M_Z  &   M_Z &  -\mu &  0 & 0 & 0 & 0 & 0 &0 \\
    \hline
    \jvs \alpha_i \Lambda^2/M_S   & \alpha_i \Lambda^2/M_S   & \beta_i \Lambda^2/M_S
    & 0 & 0&  h^{(n)}_{ij} v_u &  0& 0 & 0 \\
    \jvs 0 & 0 & \alpha_i \Lambda^2/M_S & 0 & h^{(n)}_{ji} v_u &
    \lambda_{ij}^S M_S & 0 & \delta_{i3} \Lambda & 0 \\
    \hline 
    \jvs 0 & 0 & 0 & 0 & 0&0
    & 0 &  m_\psi  & \Lambda \\
    0 & 0 & 0 & 0 &0 & \delta_{i3} \Lambda &m_\psi  & 0 &\Lambda \\
    0 & 0 & 0 & 0 &0 & 0 &\Lambda & \Lambda & 0
  \end{array}\right) \ ,
\end{align}
where $\xi \equiv \Lambda/M_P$ and $v_u $ denotes the vev of the bosonic
component of the superfields $H_u$, and we took $f_3=1$ and $f_1=f_2=0$ for
definiteness.  The coefficients $\alpha_i$ and $\beta_i$ are non-zero for
non-vanishing vevs of the the left- and right-handed sneutrinos, respectively.

The $4 \times 4$, $6 \times 6$ and $3 \times 3$ blocks on its diagonal
correspond to the MSSM neutralinos, the left- and right-handed neutrinos and
three hidden sector fermions.  In the basis in which the neutralino and the
hidden sector mass matrices are diagonal and the left and right-handed
neutrinos decouple, the matrix reads (only leading order terms are shown)
\begin{align}
  \label{eq:M2}
  \hat M &\sim
  \left( \begin{array}{c|cc|ccc}
    \jvs \scriptstyle M_{\chi \alpha \beta} & 
    \scriptstyle \tilde \alpha_{\alpha i} \Lambda^2/M_S & 
    \scriptstyle\tilde \beta_{\alpha i} \Lambda^2/M_S & \scriptstyle 0 & \scriptstyle 0 & \scriptstyle 0 \\
    \hline
    \jvs \scriptstyle\tilde \alpha_{i \alpha } \Lambda^2/M_S & 
    \scriptstyle M_{\nu_L i j}&  \scriptstyle 0 &   
   \scriptstyle -h_{i3}^{(n)} v_u \Lambda/\sqrt 2 M_3 &
   \scriptstyle h_{i3}^{(n)} v_u \Lambda/2 M_3  &
   \scriptstyle h_{i3}^{(n)} v_u \Lambda/2 M_3  \\
    \jvs  
    \scriptstyle\tilde \beta_{i \alpha } \Lambda^2/M_S
    & \scriptstyle 0 &  
    \scriptstyle M_{\nu_R i j} & 
    \scriptstyle \delta_{i3} v_u/\sqrt 2&- \delta_{i3}  v_u /2& 
   \scriptstyle- \delta_{i3}  v_u/2 \\
    \hline 
    \jvs  \scriptstyle 0 & 
    \scriptstyle -h_{3i}^{(n)} v_u \Lambda/\sqrt 2M_3   &
    \scriptstyle \delta_{3i}  v_u/\sqrt 2
    & 
    \scriptstyle -m_\psi &  \scriptstyle 0  &\scriptstyle 0 \\
    \scriptstyle 0 &
    \scriptstyle h_{3i}^{(n)} v_u \Lambda/2 M_3  &
    \scriptstyle -\delta_{3i}  v_u/2 &\scriptstyle 0  &
    \scriptstyle - \sqrt 2 \Lambda &\scriptstyle 0 \\
    \jvs \scriptstyle 0 &
    \scriptstyle h_{3i}^{(n)} v_u \Lambda/2 M_3  &
    \scriptstyle -\delta_{3i}  v_u/2 & \scriptstyle 0 &\scriptstyle 0  & 
    \scriptstyle \sqrt 2 \Lambda
  \end{array}\right) ,
\end{align}
with neutrino mass matrices $M_{\nu_L i j} =- v_u^2 \sum_k
h_{ik}^{(n)}h_{kj}^{(n)  T}/M_k$, $M_{\nu_R i j} = M_i \delta_{ij}+\mathcal
O\left(v_u^2/M_S\right) $, and diagonal neutralino masses $M_{\chi \alpha
\beta}=m_{\chi_\alpha} \delta_{\alpha \beta}$, $\alpha,\beta=1, \ldots, 4$.
\medskip

We deduce that the left-handed neutrinos mix  with lightest neutralino
$\chi_1$ and the light pseudo NG fermion $\Psi$ as
\begin{align}
  \theta_{\nu_L \chi_1} &\sim \frac{\Lambda^2}{M_S m_{\chi_1}}
  \sim 5 \times 10^{-7}
  \left( \frac{\Lambda}{10^6 \GeV} \right)^2 \left(\frac{M_S}{10^{16} \GeV}\right)^{-1}
  \left(\frac{m_{\chi_1}}{200 \GeV}\right)^{-1} \ , \\
  \theta_{\nu_{L i} \Psi} &\simeq \frac{h_{i3}^{(n)}}{\sqrt 2}\frac{\Lambda
  v_u}{M_S m_\psi}
  \simeq  1.2 \times10^{-12} \left( \frac{h_{i3}^{(n)}}{10^{-2} } \right) 
  \left( \frac{\Lambda}{10^6 \GeV} \right) \left(\frac{M_S}{10^{16} \GeV}\right)^{-1}
  \left(\frac{m_\psi}{100 \GeV}\right)^{-1} \ .
\end{align}

\section{Pseudo NG mode masses}
\label{app:Xvev}
We will shortly discuss how the pseudo NG modes of $B-L$ breaking in the
hidden sector acquire masses after SUSY breaking. To this end we will assume
that SUSY breaking is only mediated by SUGRA effects to the matter-parity
breaking sector, and that the Kahler potential is canonical.

The mass-generation of the fermionic and bosonic partner of the pseudo NG
boson, $\psi$ and $a$, is due to SUSY breaking and can be understood when
considering the hidden sector alone, neglecting couplings to the visible
sector. After breaking of SUSY, soft terms are induced in the scalar potential
of the hidden sector. They have the form
\begin{equation}
  V_\text{soft}=m_{3/2}^2
  \left( |X|^2+|V_{12}|^2+|V_{34}|^2 \right)
  +2 m_{3/2} (\Lambda^2 X + \text{h.c.})\;,
  \label{eqn:Vsoft}
\end{equation}
where we suppressed $\mathcal{O}(1)$ prefactors.\footnote{There can also be
terms of the form $m' X(V_{12}V_{34}-\Lambda^2)$, with $m'\sim m_{3/2}$, which
we will neglect.} The supersymmetric part of the scalar potential reads
\begin{equation}
  V_\text{SUSY}=|V_{12}V_{34}-\Lambda^2|^2
  +|X V_{12}|^2 + |X V_{34}|^2\;,
  \label{eqn:Vsusy}
\end{equation}
and \eqref{eqn:Vsoft} and \eqref{eqn:Vsusy} have together a minimum
at\footnote{Note, that one has to take into account additional terms of the
form $|X|^4/\Lambda^2$ in the Kahler potential to understand the stability of
the above minimum when SUSY is broken.}
\begin{align}
  \langle X\rangle &\simeq-
  m_{3/2}\left( 1+\frac{3}{2}\frac{m_{3/2}^2}{\Lambda^2} \right)\;, &
  \langle V_{12} \rangle = \langle V_{34} \rangle &\simeq 
  \Lambda - \frac{m_{3/2}^2}{\Lambda}\;.
  \label{eqn:VEVs}
\end{align}
The vev of $X$ induces a Dirac-mass term for the fermion partners of $V_{12}$
and $V_{34}$, which finally gives a mass close to $m_{3/2}$ to the fermionic
partner of the pseudo NG boson, $\psi$. By expanding the fields like
$V_{12}=\langle V_{12} \rangle + \rho/2 + \dots$ and $V_{34}=\langle V_{34}
\rangle -\rho/2+\dots$ one can furthermore derive that the mass of the bosonic
partner of the pseudo NG boson is given by $m_\rho \simeq 4 m_{3/2}$. For
definiteness, we use these masses in the phenomenological analysis, and assume
that $\psi$ is somewhat heavier than the gravitino $\psi_{3/2}$, although the
exact values can change depending on the details of the Kahler potential and
SUSY breaking sector.

Without coupling to the visible sector, the pseudo NG boson in the hidden
sector would remain massless after SUSY breaking. However, the coupling to the
right-handed neutrinos generates an additional term $(f_3^2 \Lambda^2/M_3)
V_{34}V_{34}$ in the superpotential, which after SUSY breaking give rise to
terms like $(f_3^2 m_{3/2}\Lambda^2/M_3) V_{34}V_{34} + h.c.$ in the scalar
potential. Noting that $V_{34} = \Lambda e^{-ia/2\Lambda}+\dots$, it follows
that the mass of the pseudo NG boson in the hidden sector is of the order of
$m_a\sim f_3\sqrt{m_{3/2}\Lambda^2/M_3}$, see Eq.~\eqref{eqn:NGmass}.

\end{document}